\documentclass[twocolumn,aps,amssymb,floatfix,superscriptaddress]{revtex4}

\usepackage{graphicx}
\usepackage{bm}
\usepackage{amsmath}
\usepackage{amssymb}
\usepackage{amsfonts}
\usepackage{float}
\usepackage{hyperref}
\usepackage{dsfont}  
\usepackage{slashed}  
\usepackage{booktabs}
\usepackage{multirow}
\usepackage{subfigure}
\usepackage[sort&compress]{natbib}

\newcommand{\be}{\begin{equation}}  
\newcommand{\ee}{\end{equation}}  
\newcommand{\beq}{\begin{eqnarray}}  
\newcommand{\eeq}{\end{eqnarray}}

\newcommand{\bea}{\begin{eqnarray}}
\newcommand{\eea}{\end{eqnarray}}

\begin{document}
\title{Reconstruction of light-cone parton distribution functions\\ from lattice QCD simulations at the physical point}
\author{Constantia Alexandrou}
\affiliation{
  Computation-based Science and Technology Research Center,
  The Cyprus Institute,
  20 Kavafi Street,
  Nicosia 2121,
  Cyprus}
\affiliation{
Department of Physics,
  University of Cyprus,
  P.O. Box 20537,
  1678 Nicosia,
  Cyprus}
\author{Krzysztof Cichy}
\affiliation{Faculty of Physics, Adam Mickiewicz University, Umultowska 85, 61-614 Pozna\'{n}, Poland}
\author{Martha Constantinou}
\affiliation{Department of Physics,  Temple University,  Philadelphia,  PA 19122 - 1801,  USA}
\author{Karl Jansen}
\affiliation{NIC, DESY,
  Platanenallee 6,
  D-15738 Zeuthen,
  Germany}
\author{Aurora Scapellato}
\affiliation{
  Computation-based Science and Technology Research Center,
  The Cyprus Institute,
  20 Kavafi Street,
  Nicosia 2121,
  Cyprus}
\affiliation{University of Wuppertal, Gau\ss str. 20, 42119 Wuppertal, Germany}
\author{Fernanda Steffens}
\affiliation{Institut f\"ur Strahlen- und Kernphysik, Rheinische
  Friedrich-Wilhelms-Universit\"at Bonn, Nussallee 14-16, 53115 Bonn}

\begin{abstract}

  \noindent We present the unpolarized and helicity parton distribution functions calculated within lattice QCD simulations using physical values of the light quark mass.
Non-perturbative renormalization is employed and the lattice data are converted to the $\overline{\rm MS}$-scheme at a scale of 2 GeV. 
A matching process is applied together with target mass corrections leading to the reconstruction of light-cone parton distribution functions. 
For both cases we find a similar behavior between the lattice and phenomenological data, and for the polarized PDF a nice overlap for a range of Bjorken-$x$ values. 
This presents a major success for the emerging field of direct calculations of quark distributions using lattice QCD.

\end{abstract}
\maketitle 
\bibliographystyle{apsrev}

\noindent\textit{Introduction:} 
Parton distribution functions (PDFs) are important tools that provide information on the internal dynamics of quarks and 
gluons within a hadron. Given their importance, PDFs have been a major part of both theoretical and experimental 
investigations over the last decades. Main sources of information are global QCD analyses, which 
provide accurate results due to theoretical advances and the new data emerging from accelerators, see Ref.~\cite{Lin:2017snn} for a community review. 
Despite this progress, parametrizations of PDFs are not without ambiguities~\cite{Jimenez-Delgado:2013sma}, as there are kinematical 
regions not easily accessible experimentally, e.g.\ the large Bjorken-$x$ region. 
The transversity PDF is, in addition, one example of a distribution only poorly constrained by phenomenology. 
Thus, a calculation of PDFs from first principles is of crucial importance for the deeper understanding of the inner structure of 
hadrons. It may also serve as input for experimental analysis in collision experiments, e.g.\ at LHC. Their non-perturbative nature makes lattice QCD an ideal {\textit{ab initio}} formulation to determine them, utilizing large scale simulations.  

PDFs are defined on the light cone, which poses a problem for the standard Euclidean formulation, and until recently they 
were only accessed via Mellin moments and nucleon form factors (see
e.g.~\cite{Constantinou:2014tga,Constantinou:2015agp,Alexandrou:2015yqa,Alexandrou:2015xts,Syritsyn:2014saa}
for recent studies). However, there are severe limitations in reconstruction of PDFs, mainly due to increasing 
statistical noise for high moments, and the power divergent mixing with lower dimensional operators.
Consequently, extracting PDFs from their moments is practically unfeasible.

A novel direct approach was suggested by Ji~\cite{Ji:2013dva}, who proposed a computation of spatial correlation 
functions between two boosted nucleon states, using non-local fermionic operators with a finite-length Wilson line (WL). Upon Fourier 
transform, these matrix elements (MEs) lead to so-called quasi-PDFs.
In the infinite momentum limit, contact with light-cone PDFs is reestablished via a matching 
procedure~\cite{Xiong:2013bka,Chen:2016fxx,Wang:2017qyg,Stewart:2017tvs,Izubuchi:2018srq}. 
This approach has been explored in lattice QCD with promising first results~\cite{Lin:2014zya,Alexandrou:2015rja,Chen:2016utp,Alexandrou:2016jqi}. 
Many aspects of extracting the light-cone PDFs from quasi-PDFs have improved recently. These include investigations of renormalizability~\cite{Ishikawa:2017faj}, 
development of a renormalization scheme for lattice WL operators~\cite{Alexandrou:2017huk}, refining 
the matching procedure~\cite{Wang:2017qyg,Stewart:2017tvs,Izubuchi:2018srq} and target mass corrections (TMCs)~\cite{Chen:2016utp}.
Another direct approach, proposed by Radyushkin, is pseudo-PDFs~\cite{Radyushkin:2017cyf}, a generalization of light-cone PDFs to finite 
nucleon momenta which has also been implemented in lattice QCD~\cite{Radyushkin:2017cyf,Orginos:2017kos}. Certain properties of quasi-PDFs, 
like the nucleon mass dependence and target mass effects, have also been analyzed via their relation with transverse momentum 
dependent distribution functions~\cite{Radyushkin:2017ffo,Radyushkin:2016hsy}. More recently, Ma and Qiu  proposed 
construction of lattice cross sections to study partonic structure from lattice QCD~\cite{Ma:2017pxb}. 
In this work, we present results using the approach proposed by Ji and refinements thereafter.

{\it Quasi-PDFs:} The Minkowski definition of PDFs within a hadron can be derived from the operator product expansion 
of hadronic deep inelastic scattering and is light-cone dominated, i.e.\ it receives contributions only in the region $\xi^2 {=} t^2 {-} \vec{r}^2 {\approx} 0$. 
Hence, it corresponds to a single point on the lattice, which makes it impossible to evaluate the integral that defines the PDF. Quasi-PDFs  are given by
\begin{equation}
\label{eq:quasi_pdf}
\tilde{q}(x,\Lambda,P){=}\hspace*{-0.1cm}\int_{-\infty}^{+\infty}\hspace*{-0.1cm}\frac{dz}{4\pi}\,
e^{-ixPz}\,h_\Gamma(P,z) ,
\end{equation}
where $h_\Gamma(P,z)\,=\,\langle P\vert \, \overline{\psi}(0,z)\,\Gamma W(z)\,\psi(0,0)\,\vert P\rangle$, $\Lambda{\sim} 1/a$ 
is a UV cut-off, $\vert P\rangle$ is the proton state with finite momentum $P$, which is non-zero only in the direction of the 
WL ($P{=}(P_0,0,0,P_3)$), and $z$ is the length of the WL $W(z)$ between quark fields, which is taken in a purely spatial 
direction instead of the $+$-direction on the light cone. The Dirac structure $\Gamma$ defines the type of PDF 
($\Gamma{=}\gamma_\mu$ -- unpolarized, $\Gamma{=}\gamma_5\gamma_\mu$ -- polarized and $\Gamma{=}\sigma_{\mu\nu}$ -- transversity) 
and may be taken parallel or perpendicular to the WL to avoid finite mixing (for certain lattice discretizations) with other operators~\cite{Constantinou:2017sej}. 
To account for the finite momentum used in lattice QCD simulations, higher twist corrections and TMCs need to be applied. 
For large nucleon momenta, quasi-PDFs can be matched to physical PDFs using Large Momentum Effective Theory 
(LaMET)~\cite{Ji:2013dva,Ji:2014gla}.

\noindent\textit{Lattice QCD evaluation:}
The results presented in this work are obtained with a gauge ensemble of two degenerate 
light quarks $(N_f{=} 2)$ at maximal twist, with quark masses that are tuned to reproduce approximately the physical pion mass value~\cite{Abdel-Rehim:2015pwa}. 
The parameter values of the ensemble are given in
Table~\ref{Table:params}\,. The gauge configurations have been generated with the Iwasaki improved gluon action
\cite{Iwasaki:2011np,Abdel-Rehim:2013yaa} and the twisted mass fermion action with clover improvement
\cite{Frezzotti:2003ni,Sheikholeslami:1985ij}. 
\begin{table}[h]
\begin{center}
\renewcommand{\arraystretch}{1.2}
\renewcommand{\tabcolsep}{5.5pt}
\begin{tabular}{c|lc}
\hline
\hline
\multicolumn{3}{c}{ \quad$\beta{=}2.10$,\qquad $c_{\rm SW}{=} 1.57751$,\qquad $a{=}0.0938(3)(2)$~fm\quad}\\
\hline
$48^3\times 96$\,  & $\,\,a\mu = 0.0009\quad m_N = 0.932(4)$~GeV   \\
$L=4.5$~fm\,  & $\,\,m_\pi = 0.1304(4)$~GeV$\quad m_\pi L = 2.98(1)$    \\
\hline\hline
\end{tabular}
\caption{\small{Simulation parameters of the ensemble used in this work. The nucleon mass $(m_N)$, the pion mass $(m_\pi)$ and the lattice spacing $(a)$ have been determined in Ref.~\cite{Alexandrou:2017xwd}.}}
\label{Table:params}
\end{center}
\end{table} 

High nucleon momenta are required for carrying out the matching within perturbation theory. However, 
the noise-to-signal ratio increases rapidly as the  momentum is increased, demanding a huge 
computational effort for reaching a satisfactory  statistical accuracy. There are additional factors that 
contribute to the increase of gauge noise, such as using the physical pion mass
and large enough propagation in Euclidean time to suppress excited states.

In this study, we compute quasi-PDFs for three values of the momentum, namely $\frac{6\pi}{L}$, $\frac{8\pi}{L}$ and $\frac{10\pi}{L}$, which in physical units 
correspond to $0.83,\,1.11,\,1.38$~GeV. We implement the momentum smearing technique~\cite{Bali:2016lva}, which is necessary to achieve high momentum 
at a reasonable computational cost~\cite{Alexandrou:2016jqi}. The total number of measurements
for momenta $\frac{6\pi}{L}$, $\frac{8\pi}{L}$, $\frac{10\pi}{L}$ is 4800-9600, 38250, 58950, respectively. 
Going to even larger momentum, although  desirable, requires 
huge computational resources.
 
In the computation of MEs, we apply up to 20 iterations of stout smearing~\cite{Morningstar:2003gk} to gauge links of the operator.
This reduces the power divergence in the ME of non-local bilinear operators connected with a WL and  brings renormalization functions 
(Z-factors) closer to their tree level value. After carrying out the power divergence subtraction, renormalized MEs extracted from different stout levels
must be in agreement. This provides a check of the renormalization process. 

\begin{figure}[h]
\centering
\includegraphics[scale=0.31]{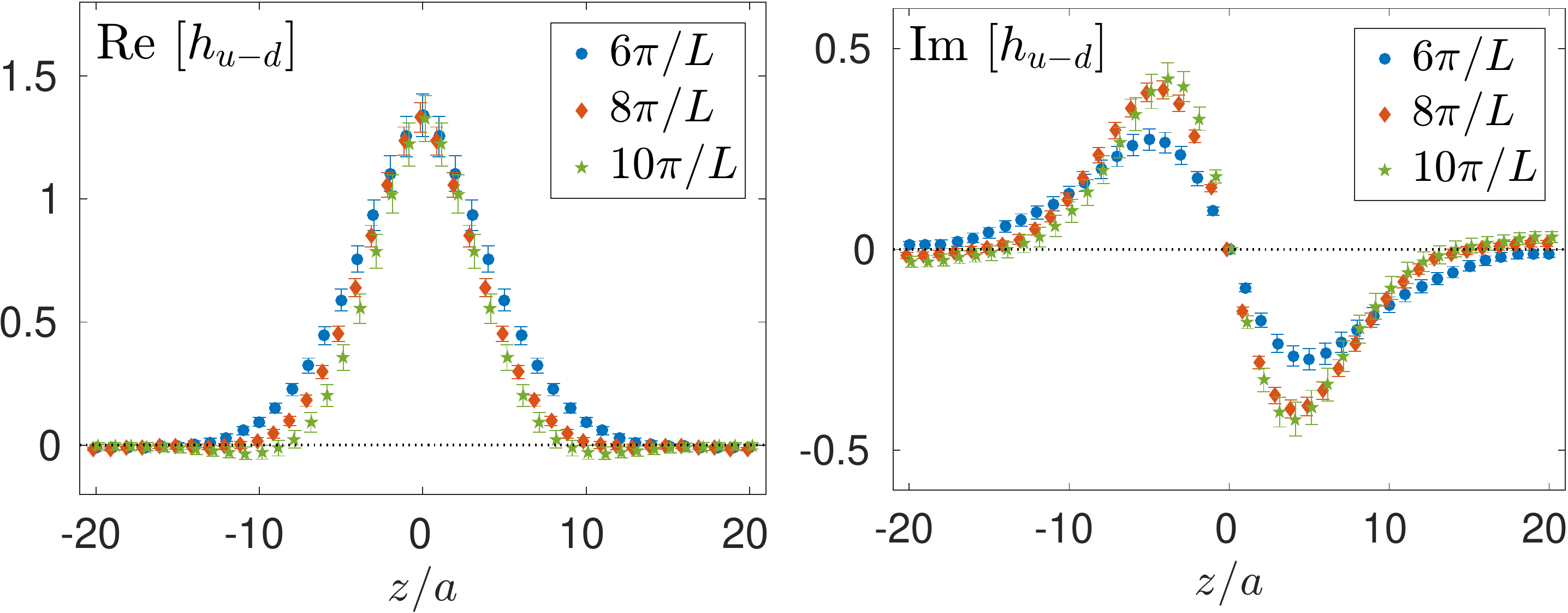}
\vskip -0.25cm
\caption{\small{Comparison of unpolarized bare MEs ($h_{\gamma_0}$) for momenta $\frac{6\pi}{L}$ (blue circles), $\frac{8\pi}{L}$ (red diamonds) 
and $\frac{10\pi}{L}$ (green stars) using 5 stout steps.}}
\label{fig:g0_g3}
\end{figure}

\begin{figure}[h]
\centering
\includegraphics[scale=0.31]{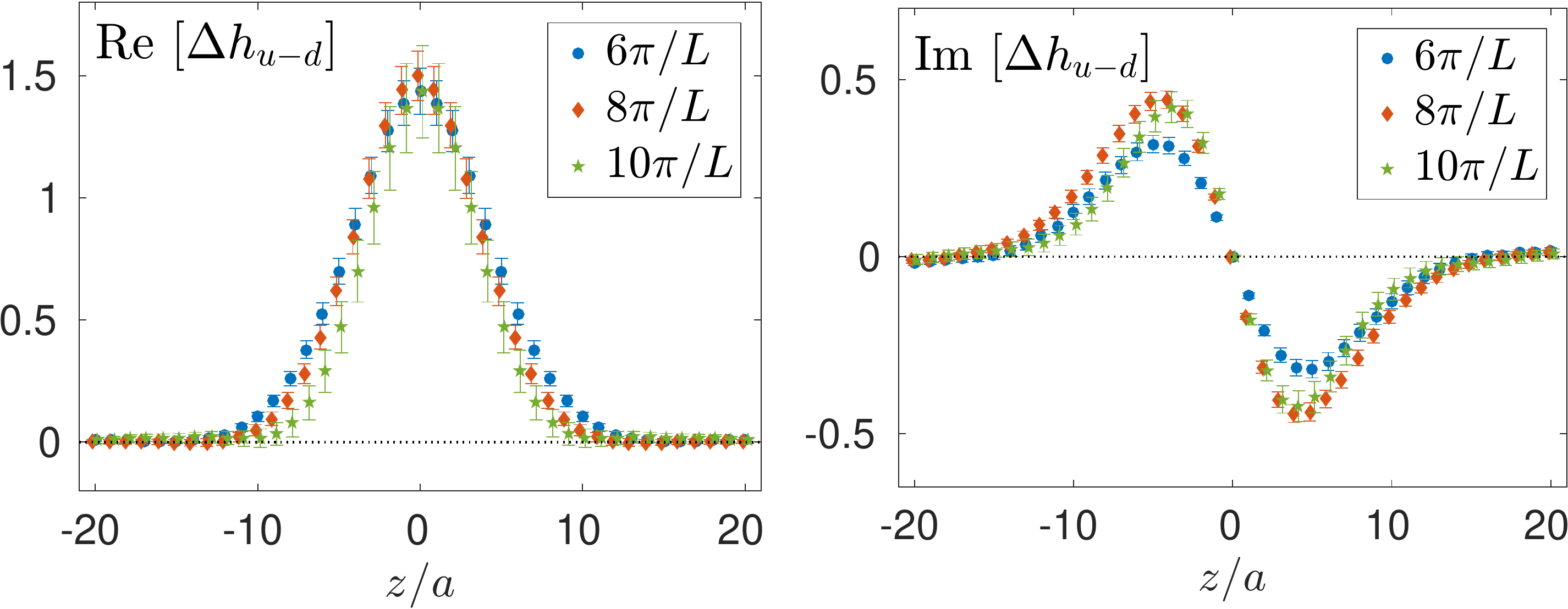}
\vskip -0.25cm
\caption{\small{Similar to Fig.\ref{fig:g0_g3} for helicity bare MEs.}}
\label{fig:bare_g5g3}
\end{figure}

As mentioned above, one can extract the unpolarized PDF from an operator with a Dirac structure parallel ($\gamma_3$) or perpendicular to the WL 
($\gamma_0$). The former has the disadvantage of mixing with the twist-3 scalar operator~\cite{Constantinou:2017sej}. However, for twisted mass fermions 
the vector mixes with the pseudoscalar operator, which vanishes in the continuum limit. As a consequence, $h_{\gamma_3}$ has increased noise 
contamination compared to $h_{\gamma_0}$. We compute MEs of both operators, and here we focus on $h_{\gamma_0}$ presented in Fig.~\ref{fig:g0_g3}
for the three momenta values momenta used. Similarly, in Fig.~\ref{fig:bare_g5g3}, we show results for bare helicity MEs. 

\vspace{0.1cm}

It is evident that the signal quality rapidly worsens for larger momenta, and an increase in  statistics by a factor four to six  is used for momenta 
$\frac{8\pi}{L}$ and $\frac{10\pi}{L}$ as compared to $\frac{6\pi}{L}$, to keep statistical uncertainties under control. 
As can be seen from Figs.~\ref{fig:g0_g3}-\ref{fig:bare_g5g3}, results for the two largest momentum values are overlapping for both the real 
and imaginary parts within our statistical errors.

\noindent\textit{Renormalization:} 
To obtain physical results, lattice MEs of non-conserved currents must be renormalized to eliminate divergences. Compared to other nucleon quantities, 
quasi-PDFs have an additional WL-related power divergence. Based on the renormalization and mixing pattern from Ref.~\cite{Constantinou:2017sej},
we developed a non-perturbative prescription~\cite{Alexandrou:2017huk}, also implemented for another lattice formulation~\cite{Chen:2017mzz}. 
This procedure removes the power divergence and the logarithmic divergence with respect to the regulator, and applies the necessary finite renormalization related to the 
lattice regularization. 
For our choices of the Dirac structure for the unpolarized and the polarized cases, there is no mixing.

We adopt a non-perturbatively applicable RI$'$-type scheme~\cite{Martinelli:1994ty} and Z-factors are extracted by imposing conditions
described in Ref.~\cite{Alexandrou:2015sea}. The RI$'$ renormalization scale is chosen to be of the form $(n_t,n,n,n)$ in order to suppress 
discretization effects~\cite{Alexandrou:2015sea}. An extensive study on the choice for the renormalization scale can be found in 
Ref.~\cite{Alexandrou:2017huk}. In this work, we use a total of 17 choices for the RI$'$ scale that covers the range $(a\,p)^2 \in [0.7,2.6]$. 
The Z-factors are converted to the $\overline{\rm MS}$ scheme and evolved to $\mu{=}2$ GeV using the formulae of Ref.~\cite{Constantinou:2017sej}. 
Final estimates are extracted from the extrapolation $(a\,p)^2 {\to} 0$ using a linear fit and data in the region $[1,2.6]$. 
 
\begin{figure}[h]
\centering
\includegraphics[scale=0.31]{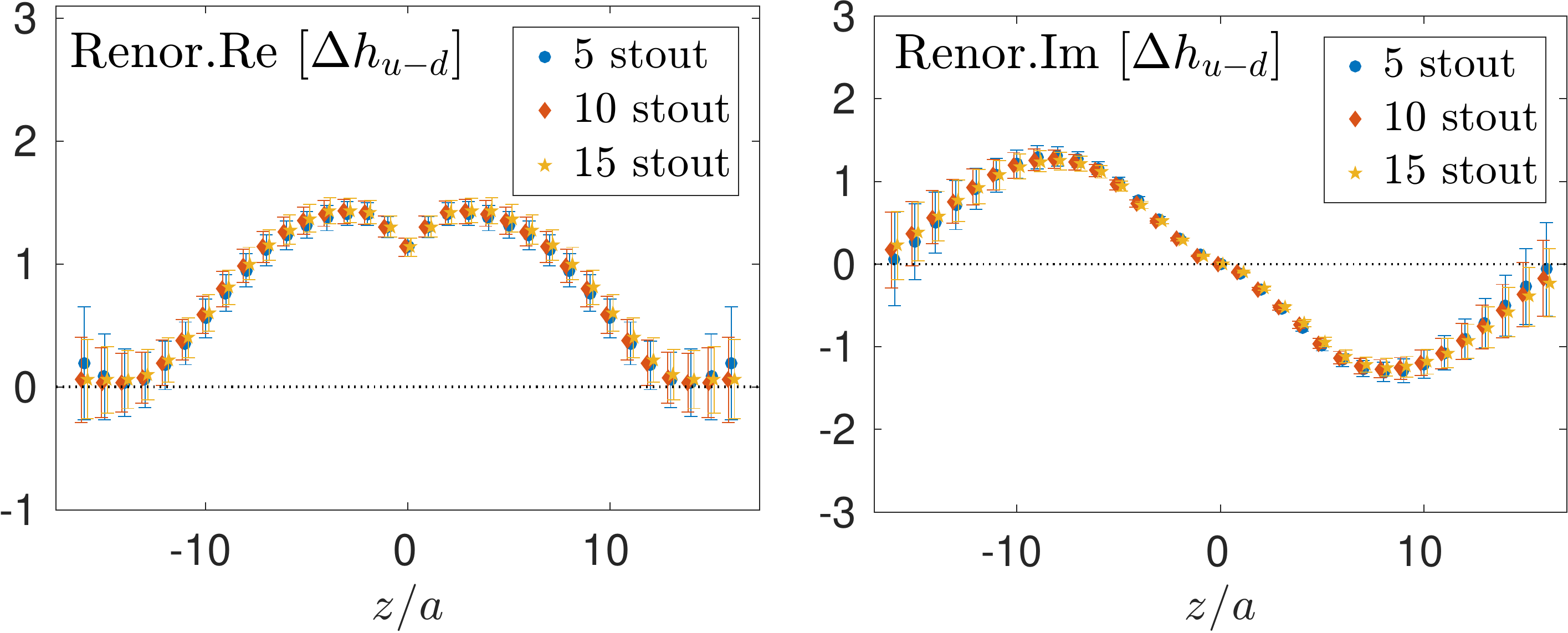}
\vskip -0.25cm
\caption{\small{Real (left) and imaginary (right) part of renormalized helicity MEs for momentum $\frac{6\pi}{L}$, as a function 
of the WL length. Blue circles/red diamonds/orange stars correspond to 5/10/15 iterations of stout smearing.}}
\label{fig:helicity_Renorm}
\end{figure} 
For demonstration purposes, we present renormalized helicity MEs for momentum $\frac{6\pi}{L}$ in Fig.~\ref{fig:helicity_Renorm}. 
Renormalization functions and MEs share similar properties with respect to $z$ (symmetric real part and antisymmetric imaginary part).
As already pointed out, the stout smearing modifies both the values of MEs and Z-factors, but upon renormalization this dependence 
should vanish. In Fig.~\ref{fig:helicity_Renorm}, we compare renormalized helicity MEs extracted using 5, 10 and 15 stout smearing steps, in the $\overline{\rm MS}$ scheme at
2 GeV. The agreement found between the three cases validates the success of the renormalization prescription.
It is worth mentioning that the agreement is more prominent upon the $(a\,p)^2 {\to} 0$ extrapolation of Z-factors and holds also for the unpolarized case.

\noindent\textit{Matching to light-cone PDFs:}
Quasi-PDFs are extracted from the Fourier transform of renormalized MEs.
To obtain the light-cone PDF from quasi-PDF, one needs to apply a pertubative matching
procedure~\cite{Xiong:2013bka,Alexandrou:2015rja,Chen:2016fxx,Wang:2017qyg,Stewart:2017tvs,Izubuchi:2018srq}, 
valid thanks to the fact that infrared physics is the same for both quasi and light-cone PDFs.
The matching formula can be expressed as
\begin{eqnarray}
\label{eq:matching}
q ( x,\mu )
  =\! \int_{-\infty}^\infty \frac{d\xi}{|\xi|}\, C \left(
   \xi, \frac{\mu}{xP_3} \right) \tilde{q} \left(\frac{x}{\xi},\mu,P_3\right)\!,
\end{eqnarray}
where $\tilde{q}\left(x,\mu,P_3\right)$ is the renormalized quasi-PDF and $q(x,\mu)$ is the light-cone (matched) renormalized PDF.
$C$ represents the matching kernel and here we use a modified expression of the one suggested in Ref.~\cite{Izubuchi:2018srq}, given by
\begin{widetext}
\begin{eqnarray}
\label{eq:kernel}
C\left( \xi, \frac{\xi\mu}{xP_3} \right)=\delta(1-\xi)+\frac{\alpha_s}{2\pi}\,C_F\,\left\{
\begin{array}{ll}
\displaystyle \left[\frac{1+\xi^2}{1-\xi}\ln\frac{\xi}{\xi-1} + 1 + {3\over 2\xi}\right]_+
&\, \xi>1,
\\[10pt]
\displaystyle \left[\frac{1+\xi^2}{1-\xi}
\ln\frac{x^2P_3^2}{\xi^2\mu^2}\left(4\xi(1-\xi)\right) - \frac{\xi(1+\xi)}{1-\xi}+2\iota(1-\xi)  \right]_+
&\, 0<\xi<1,
\\[10pt]
\displaystyle \left[-\frac{1+\xi^2}{1-\xi}\ln\frac{\xi}{\xi-1} - 1 + \frac{3}{2(1-\xi)}\right]_+
&\, \xi<0,
\end{array}\right.
\end{eqnarray}
\end{widetext}
to one-loop order. In (\ref{eq:kernel}), $\iota{=}0$ for $\gamma_0$ and $\iota{=}1$ for $\gamma_3$/$\gamma_5\gamma_3$.
The plus prescriptions in the above equation are all at $\xi{=}1$.
Unlike the light-cone PDF case, for quasi-PDF a UV divergence in the one-loop wave function correction appears as an integral in the momentum fraction, which
can be regularized using standard dimensional regularization. The wave function renormalization is then computed in the usual way, and it is expressed by
integrals of $-3/2\xi$ ($\xi{>}1$) and $-3/2(1{-}\xi)$ ($\xi{<}0$). From a Ward identity, we have that the integrated one-loop vertex correction is renormalized 
by the same terms. This ensures that the normalization of the distributions is automatically preserved by the matching, that is, from Eqs. (\ref{eq:matching}) and (\ref{eq:kernel}), 
one has $\int_{-\infty}^{\infty} dx\, q(x,\mu) {=} \int_{-\infty}^{\infty} dx\, \tilde{q}(x,\mu,P_3)$, and $\int_{-\infty}^{\infty} d\xi\, C(\xi,\xi\mu/xP_3){=}1$. 
The final step after the matching is to apply TMCs according to formulae of Ref.~\cite{Chen:2016utp}.
A complete discussion about the matching and its comparison with other studies~\cite{Wang:2017qyg,Stewart:2017tvs,Izubuchi:2018srq} will be presented in a separate publication.

\noindent\textit{Results:} 
The ultimate goal of this work is the extraction the Bjorken-$x$ dependence from lattice QCD. This is achieved by computing MEs 
for WL operators and applying the renormalization followed by Fourier transform to extract quasi-PDFs. Finally, the matching 
procedure and TMCs are applied to reconstruct light-cone PDFs.

\begin{figure}[h]
\centering
\includegraphics[scale=0.52]{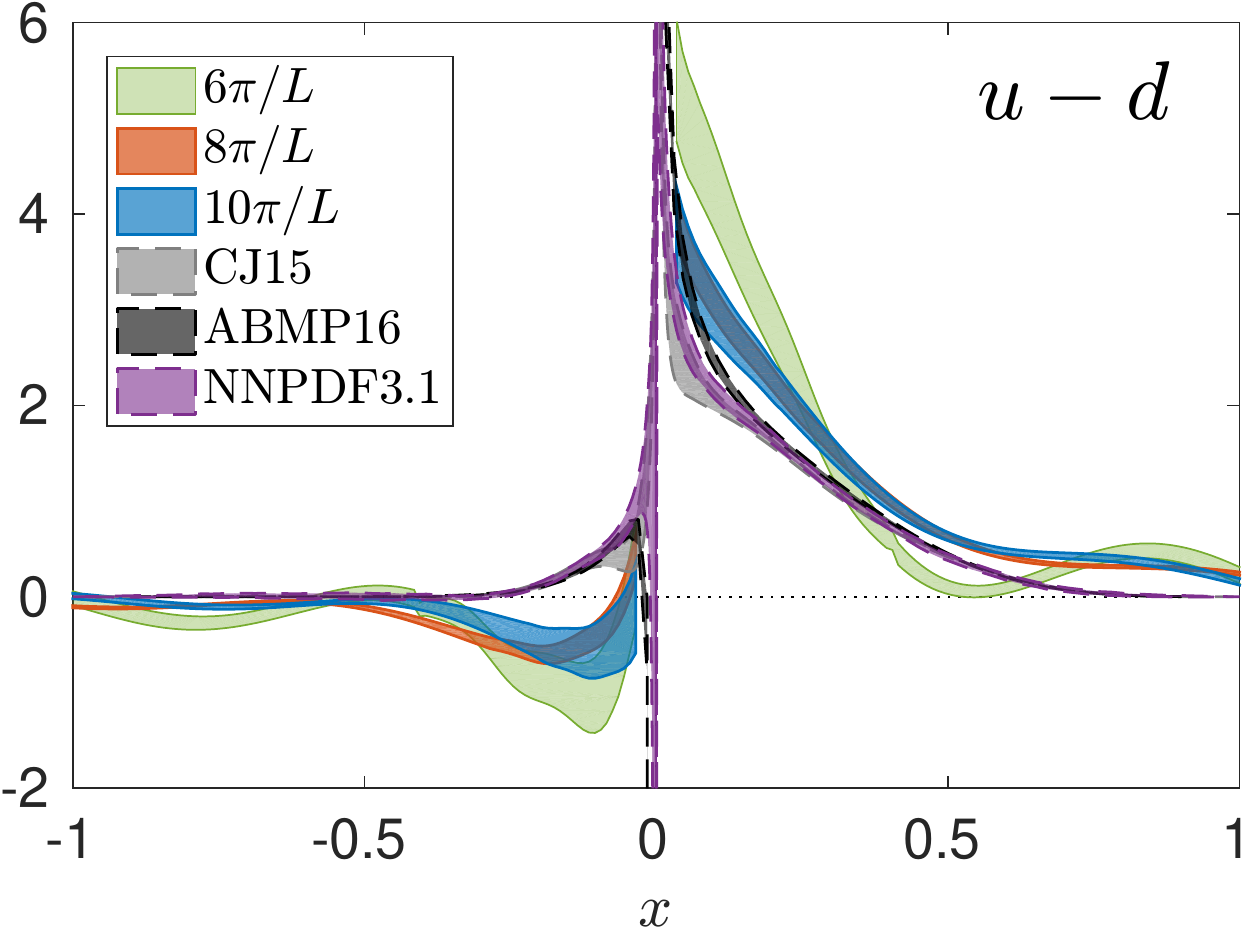}
\vspace*{-0.35cm}
\caption{\small{Comparison of unpolarized PDF
at momenta $\frac{6\pi}{L}$ (green band), $\frac{8\pi}{L}$ (orange band), 
$\frac{10\pi}{L}$ (blue band), and ABMP16~\cite{Alekhin:2017kpj} (NNLO), 
NNPDF~\cite{Ball:2017nwa} (NNLO) and CJ15~\cite{Accardi:2016qay} (NLO) phenomenological curves.}}
\label{fig:matched_unpol}
\end{figure}

In Fig.~\ref{fig:matched_unpol}, we show our final results for the unpolarized PDF using the temporal direction for
the Dirac structure ($h_{\gamma_0}$), which is free of mixing and leads to higher statistical accuracy than $h_{\gamma_3}$.
We show the  dependence on the nucleon momentum for values $\frac{6\pi}{L}$,  $\frac{8\pi}{L}$ and $\frac{10\pi}{L}$ as well as
 the phenomenological determinations  CJ15~\cite{Accardi:2016qay}, ABMP16~\cite{Alekhin:2017kpj} and NNPDF31~\cite{Ball:2017nwa}.
We find that as the momentum increases, the data approach phenomenological results. In particular, increasing the nucleon 
momentum from $\frac{6\pi}{L}$ to $\frac{8\pi}{L}$ has a large effect on the PDFs shape, with the latter approaching
the phenomenological curve.  Furthermore, we find a saturation of PDFs for $\frac{8\pi}{L}$ and 
$\frac{10\pi}{L}$, indicating that LaMET may be applicable for $P\geq\frac{8\pi}{L}$. 
The interplay of real and imaginary parts of renormalized MEs leads to unphysical oscillations in quasi-PDFs, resulting from the periodicity of the Fourier transform, and propagated 
through the matching procedure to light-cone PDFs. The effect is naturally suppressed for large nucleon boosts, when MEs decay to zero fast enough, before $e^{-ixPz}$ becomes 
negative. For the currently attained momenta, the decay of renormalized ME is still relatively slow (cf. Fig.~\ref{fig:helicity_Renorm}), which manifests itself in distorted approach of 
the PDF to zero for $x\gtrsim0.5$ and unphysical minimum in the antiquark part, for $x\approx-0.2$. The oscillations, as expected, are smoothened out as the momentum increases 
(which is visible particularly at the level of quasi-PDFs), and are more severe in the negative region.
Nevertheless, this is the first time when clear convergence is  demonstrated with simulations using a physical pion mass value.
Clearly, momentum $\frac{6\pi}{L}$ is not high enough to reconstruct light-cone PDFs. 
However, we observe a similar behavior of the lattice data at momentum $\frac{10\pi}{L}$ as compared to 
phenomenological results, with some overlap in the small-$x$ region. 
The slope of the two curves is compatible for the positive-$x$ region,
and both curves go to zero for $x\lesssim-0.3$ and $x\gtrsim1$. 
Compatible results are extracted for $h_{\gamma_3}$, but with increased uncertainties. 

\begin{figure}[h!]
\centering
\includegraphics[scale=0.52]{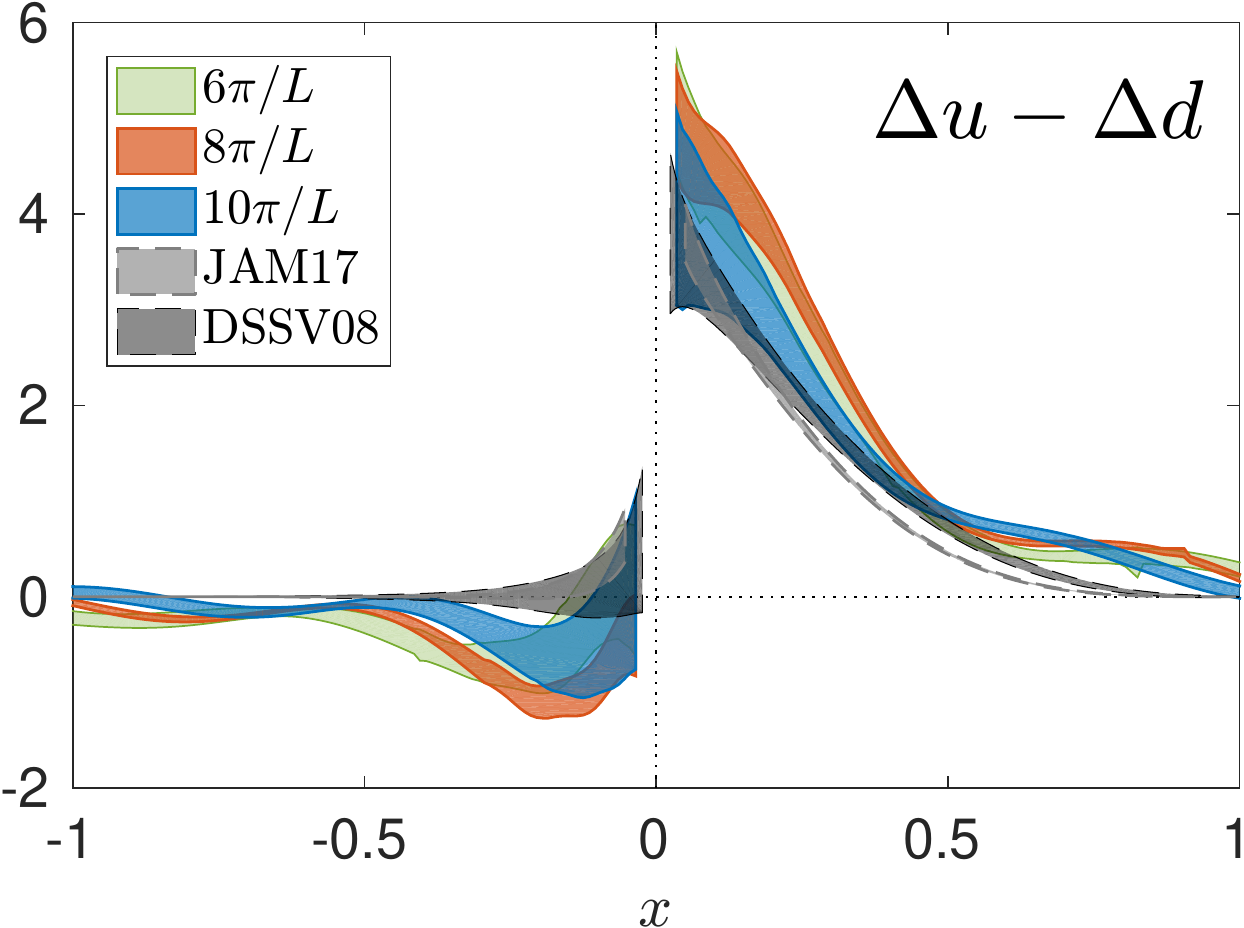}
\vspace*{-0.35cm}
\caption{\small{Comparison of polarized PDF
at momenta $\frac{6\pi}{L}$ (green band), $\frac{8\pi}{L}$ (orange band), 
$\frac{10\pi}{L}$ (blue band), DSSV08~\cite{deFlorian:2009vb} and JAM17 NLO phenomenological data~\cite{Ethier:2017zbq}.}}
\label{fig:matched_pol}
\end{figure}

In Fig.~\ref{fig:matched_pol}, we present polarized PDFs for the three momenta, together with 
DSSV08~\cite{deFlorian:2009vb} and JAM17~\cite{Ethier:2017zbq} phenomenological data.
We find a milder dependence on the nucleon momentum, and $\frac{10\pi}{L}$
is much closer to phenomenological curves with significant overlap with the JAM17 data for $0<x<0.5$.
For the region $0.5<x<1$, the slope of the lattice data changes, possibly due to oscillations mentioned above, but it approaches
zero around $x=1$. For the negative-$x$ region, the lattice data also approach zero, with a dip at small-$x$ and
large uncertainties, another consequence of oscillations. 

\begin{figure}[h!]
\centering
\includegraphics[scale=0.95]{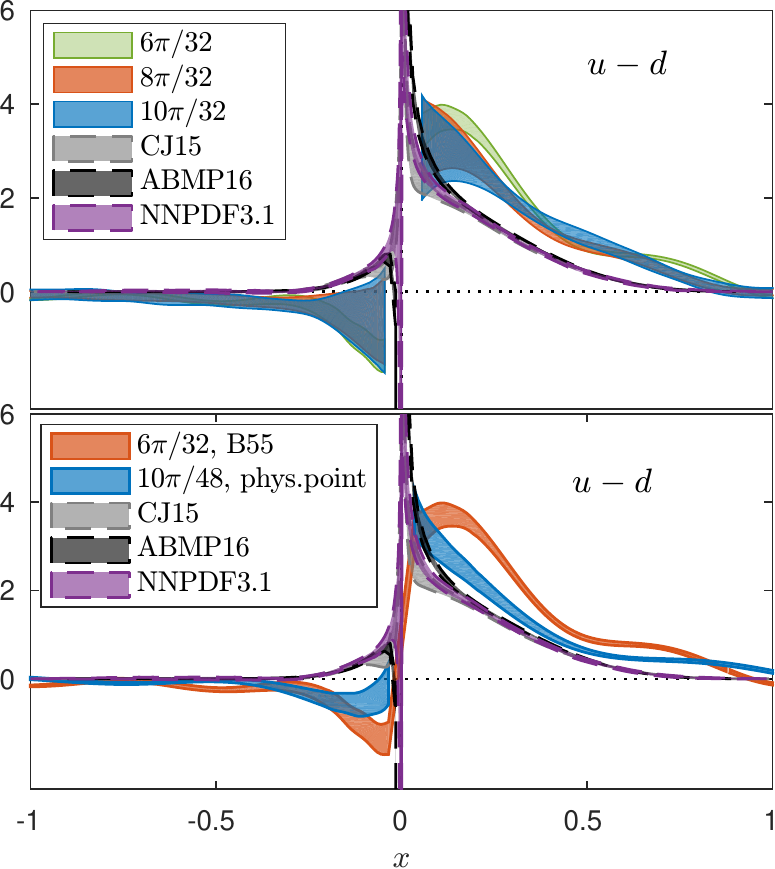}
\vskip -.35cm
\caption{\small{Top: Comparison of unpolarized PDF from the B55 ensemble against phenomenological estimates. Notation
as in Fig.~\ref{fig:matched_unpol}. 
Bottom: Comparison of unpolarized PDF between results of this work (blue band) and of the B55 ensemble (orange band) at nucleon momentum ${\sim}1.4$ GeV.}}
\label{fig:B55_vs_phys}
\end{figure}

Simulating at the physical point is crucial for obtaining data that are close to the global analyses data. This is demonstrated in Fig.~\ref{fig:B55_vs_phys}. 
In the top panel, we compare  phenomenological estimates with results from Ref.~\cite{Alexandrou:2016jqi} at $m_\pi{=}375$ MeV and volume 
$32^3{\times}64$ (B55). As the nucleon momentum increases, one observes that the B55 data saturate away from phenomenological curves, wrongfully 
leading to discouraging conclusions for quasi-PDFs approach. In the lower panel of Fig.~\ref{fig:B55_vs_phys}, we plot data from this work  with the B55 
ensemble, both at momentum ${\sim}1.4$ GeV. As can be seen, there is a clear pion mass dependence and the B55 data are away from the global analyses curves.

\noindent\textit{Conclusions}:

We present the first ever lattice calculation of unpolarized and helicity PDFs  where long-standing obstacles, 
such as large momenta, physical pion mass and non-perturbative renormalization have been addressed.
To investigate the nucleon momentum dependence, we employed three values corresponding to 
$0.83,\,1.11,\,1.38$~GeV, with appropriately increased number of measurements for the latter ones 
to keep statistical uncertainties under control.

Lattice MEs are renormalized non-perturbatively in the RI$'$ scheme and are converted to the $\overline{\rm MS}$-scheme
at $\mu{=}2$ GeV. Light-cone PDFs are reconstructed upon Fourier transform and matching with target mass corrections.
Our final results for PDFs are highlighted in Figs.~\ref{fig:matched_unpol},\ref{fig:matched_pol}.
We are able to  compare with phenomenological results for the first time, as all necessary steps of
extracting physical PDFs have been applied and no chiral extrapolation is needed. As shown in Fig.~\ref{fig:B55_vs_phys}, there is
strong pion mass dependence and a similar behavior between lattice and phenomenology is only established at the physical pion mass ensemble. 
A further investigation of possible discretization and volume effects, as well as an improved treatment of 
the unphysical oscillations, will be pursued in the near future.

\noindent\textit{Acknowledgements}:
We would like to thank all members of ETMC for their constant and pleasant collaboration. 
We also thank the CJ, ABMP and NNPDF collaborations for providing their phenomenological
parameterizations. This work has received funding from the European Union's Horizon 2020 
research and innovation programme under the Marie Sk\l{}odowska-Curie grant agreement 
No 642069 (HPC-LEAP). K.C.\ was supported by National Science Centre (Poland) grant SONATA 
BIS no.\ 2016/22/E/ST2/00013. F.S.\ was funded by DFG project number 392578569.
M.C. acknowledges financial support by the U.S. Department of Energy, Office of Nuclear Physics, within
the framework of the TMD Topical Collaboration, as well as, by the National Science Foundation
under Grant No.\ PHY-1714407. This research used computational resources provided by the Titan
supercomputer at the Oak Ridge Leadership Computing Facility (OLCF), Prometheus supercomputer 
at the Academic Computing Centre Cyfronet AGH in Cracow (grant ID \textit{quasipdfs}), Eagle 
supercomputer at the Poznan Supercomputing and Networking Center (grant no.\ 346), Okeanos 
supercomputer at the Interdisciplinary Centre for Mathematical and Computational Modelling in 
Warsaw (grant IDs gb70-17, ga71-22). 

\bibliography{references}

\begin{thebibliography}{42}
\expandafter\ifx\csname natexlab\endcsname\relax\def\natexlab#1{#1}\fi
\expandafter\ifx\csname bibnamefont\endcsname\relax
  \def\bibnamefont#1{#1}\fi
\expandafter\ifx\csname bibfnamefont\endcsname\relax
  \def\bibfnamefont#1{#1}\fi
\expandafter\ifx\csname citenamefont\endcsname\relax
  \def\citenamefont#1{#1}\fi
\expandafter\ifx\csname url\endcsname\relax
  \def\url#1{\texttt{#1}}\fi
\expandafter\ifx\csname urlprefix\endcsname\relax\def\urlprefix{URL }\fi
\providecommand{\bibinfo}[2]{#2}
\providecommand{\eprint}[2][]{\url{#2}}

\bibitem[{\citenamefont{Lin et~al.}(2017)}]{Lin:2017snn}
\bibinfo{author}{\bibfnamefont{H.-W.} \bibnamefont{Lin}} \bibnamefont{et~al.}
  (\bibinfo{year}{2017}), \eprint{1711.07916}.

\bibitem[{\citenamefont{Jimenez-Delgado
  et~al.}(2013)\citenamefont{Jimenez-Delgado, Melnitchouk, and
  Owens}}]{Jimenez-Delgado:2013sma}
\bibinfo{author}{\bibfnamefont{P.}~\bibnamefont{Jimenez-Delgado}},
  \bibinfo{author}{\bibfnamefont{W.}~\bibnamefont{Melnitchouk}},
  \bibnamefont{and} \bibinfo{author}{\bibfnamefont{J.~F.} \bibnamefont{Owens}},
  \bibinfo{journal}{J. Phys.} \textbf{\bibinfo{volume}{G40}},
  \bibinfo{pages}{093102} (\bibinfo{year}{2013}), \eprint{1306.6515}.

\bibitem[{\citenamefont{Constantinou}(2015{\natexlab{a}})}]{Constantinou:2014tga}
\bibinfo{author}{\bibfnamefont{M.}~\bibnamefont{Constantinou}},
  \bibinfo{journal}{PoS} \textbf{\bibinfo{volume}{LATTICE2014}},
  \bibinfo{pages}{001} (\bibinfo{year}{2015}{\natexlab{a}}),
  \eprint{1411.0078}.

\bibitem[{\citenamefont{Constantinou}(2015{\natexlab{b}})}]{Constantinou:2015agp}
\bibinfo{author}{\bibfnamefont{M.}~\bibnamefont{Constantinou}},
  \bibinfo{journal}{PoS} \textbf{\bibinfo{volume}{CD15}}, \bibinfo{pages}{009}
  (\bibinfo{year}{2015}{\natexlab{b}}), \eprint{1511.00214}.

\bibitem[{\citenamefont{Alexandrou and Jansen}(2015)}]{Alexandrou:2015yqa}
\bibinfo{author}{\bibfnamefont{C.}~\bibnamefont{Alexandrou}} \bibnamefont{and}
  \bibinfo{author}{\bibfnamefont{K.}~\bibnamefont{Jansen}},
  \bibinfo{journal}{Nucl. Part. Phys. Proc.}
  \textbf{\bibinfo{volume}{261-262}}, \bibinfo{pages}{202}
  (\bibinfo{year}{2015}).

\bibitem[{\citenamefont{Alexandrou}(2015)}]{Alexandrou:2015xts}
\bibinfo{author}{\bibfnamefont{C.}~\bibnamefont{Alexandrou}}, in
  \emph{\bibinfo{booktitle}{{Proceedings, 45th International Symposium on
  Multiparticle Dynamics (ISMD 2015): Kreuth, Germany, October 4-9, 2015}}}
  (\bibinfo{year}{2015}), \eprint{1512.03924}.

\bibitem[{\citenamefont{Syritsyn}(2014)}]{Syritsyn:2014saa}
\bibinfo{author}{\bibfnamefont{S.}~\bibnamefont{Syritsyn}},
  \bibinfo{journal}{PoS} \textbf{\bibinfo{volume}{LATTICE2013}},
  \bibinfo{pages}{009} (\bibinfo{year}{2014}), \eprint{1403.4686}.

\bibitem[{\citenamefont{Ji}(2013)}]{Ji:2013dva}
\bibinfo{author}{\bibfnamefont{X.}~\bibnamefont{Ji}},
  \bibinfo{journal}{Phys.Rev.Lett.} \textbf{\bibinfo{volume}{110}},
  \bibinfo{pages}{262002} (\bibinfo{year}{2013}), \eprint{1306.1539}.

\bibitem[{\citenamefont{Xiong et~al.}(2014)\citenamefont{Xiong, Ji, Zhang, and
  Zhao}}]{Xiong:2013bka}
\bibinfo{author}{\bibfnamefont{X.}~\bibnamefont{Xiong}},
  \bibinfo{author}{\bibfnamefont{X.}~\bibnamefont{Ji}},
  \bibinfo{author}{\bibfnamefont{J.-H.} \bibnamefont{Zhang}}, \bibnamefont{and}
  \bibinfo{author}{\bibfnamefont{Y.}~\bibnamefont{Zhao}},
  \bibinfo{journal}{Phys.Rev.} \textbf{\bibinfo{volume}{D90}},
  \bibinfo{pages}{014051} (\bibinfo{year}{2014}), \eprint{1310.7471}.

\bibitem[{\citenamefont{Chen et~al.}(2017)\citenamefont{Chen, Ji, and
  Zhang}}]{Chen:2016fxx}
\bibinfo{author}{\bibfnamefont{J.-W.} \bibnamefont{Chen}},
  \bibinfo{author}{\bibfnamefont{X.}~\bibnamefont{Ji}}, \bibnamefont{and}
  \bibinfo{author}{\bibfnamefont{J.-H.} \bibnamefont{Zhang}},
  \bibinfo{journal}{Nucl. Phys.} \textbf{\bibinfo{volume}{B915}},
  \bibinfo{pages}{1} (\bibinfo{year}{2017}), \eprint{1609.08102}.

\bibitem[{\citenamefont{Wang et~al.}(2017)\citenamefont{Wang, Zhao, and
  Zhu}}]{Wang:2017qyg}
\bibinfo{author}{\bibfnamefont{W.}~\bibnamefont{Wang}},
  \bibinfo{author}{\bibfnamefont{S.}~\bibnamefont{Zhao}}, \bibnamefont{and}
  \bibinfo{author}{\bibfnamefont{R.}~\bibnamefont{Zhu}} (\bibinfo{year}{2017}),
  \eprint{1708.02458}.

\bibitem[{\citenamefont{Stewart and Zhao}(2017)}]{Stewart:2017tvs}
\bibinfo{author}{\bibfnamefont{I.~W.} \bibnamefont{Stewart}} \bibnamefont{and}
  \bibinfo{author}{\bibfnamefont{Y.}~\bibnamefont{Zhao}}
  (\bibinfo{year}{2017}), \eprint{1709.04933}.

\bibitem[{\citenamefont{Izubuchi et~al.}(2018)\citenamefont{Izubuchi, Ji, Jin,
  Stewart, and Zhao}}]{Izubuchi:2018srq}
\bibinfo{author}{\bibfnamefont{T.}~\bibnamefont{Izubuchi}},
  \bibinfo{author}{\bibfnamefont{X.}~\bibnamefont{Ji}},
  \bibinfo{author}{\bibfnamefont{L.}~\bibnamefont{Jin}},
  \bibinfo{author}{\bibfnamefont{I.~W.} \bibnamefont{Stewart}},
  \bibnamefont{and} \bibinfo{author}{\bibfnamefont{Y.}~\bibnamefont{Zhao}}
  (\bibinfo{year}{2018}), \eprint{1801.03917v1}.

\bibitem[{\citenamefont{Lin et~al.}(2015)\citenamefont{Lin, Chen, Cohen, and
  Ji}}]{Lin:2014zya}
\bibinfo{author}{\bibfnamefont{H.-W.} \bibnamefont{Lin}},
  \bibinfo{author}{\bibfnamefont{J.-W.} \bibnamefont{Chen}},
  \bibinfo{author}{\bibfnamefont{S.~D.} \bibnamefont{Cohen}}, \bibnamefont{and}
  \bibinfo{author}{\bibfnamefont{X.}~\bibnamefont{Ji}}, \bibinfo{journal}{Phys.
  Rev.} \textbf{\bibinfo{volume}{D91}}, \bibinfo{pages}{054510}
  (\bibinfo{year}{2015}), \eprint{1402.1462}.

\bibitem[{\citenamefont{Alexandrou et~al.}(2015)\citenamefont{Alexandrou,
  Cichy, Drach, Garcia-Ramos, Hadjiyiannakou, Jansen, Steffens, and
  Wiese}}]{Alexandrou:2015rja}
\bibinfo{author}{\bibfnamefont{C.}~\bibnamefont{Alexandrou}},
  \bibinfo{author}{\bibfnamefont{K.}~\bibnamefont{Cichy}},
  \bibinfo{author}{\bibfnamefont{V.}~\bibnamefont{Drach}},
  \bibinfo{author}{\bibfnamefont{E.}~\bibnamefont{Garcia-Ramos}},
  \bibinfo{author}{\bibfnamefont{K.}~\bibnamefont{Hadjiyiannakou}},
  \bibinfo{author}{\bibfnamefont{K.}~\bibnamefont{Jansen}},
  \bibinfo{author}{\bibfnamefont{F.}~\bibnamefont{Steffens}}, \bibnamefont{and}
  \bibinfo{author}{\bibfnamefont{C.}~\bibnamefont{Wiese}},
  \bibinfo{journal}{Phys. Rev.} \textbf{\bibinfo{volume}{D92}},
  \bibinfo{pages}{014502} (\bibinfo{year}{2015}), \eprint{1504.07455}.

\bibitem[{\citenamefont{Chen et~al.}(2016)\citenamefont{Chen, Cohen, Ji, Lin,
  and Zhang}}]{Chen:2016utp}
\bibinfo{author}{\bibfnamefont{J.-W.} \bibnamefont{Chen}},
  \bibinfo{author}{\bibfnamefont{S.~D.} \bibnamefont{Cohen}},
  \bibinfo{author}{\bibfnamefont{X.}~\bibnamefont{Ji}},
  \bibinfo{author}{\bibfnamefont{H.-W.} \bibnamefont{Lin}}, \bibnamefont{and}
  \bibinfo{author}{\bibfnamefont{J.-H.} \bibnamefont{Zhang}},
  \bibinfo{journal}{Nucl. Phys.} \textbf{\bibinfo{volume}{B911}},
  \bibinfo{pages}{246} (\bibinfo{year}{2016}), \eprint{1603.06664}.

\bibitem[{\citenamefont{Alexandrou
  et~al.}(2017{\natexlab{a}})\citenamefont{Alexandrou, Cichy, Constantinou,
  Hadjiyiannakou, Jansen, Steffens, and Wiese}}]{Alexandrou:2016jqi}
\bibinfo{author}{\bibfnamefont{C.}~\bibnamefont{Alexandrou}},
  \bibinfo{author}{\bibfnamefont{K.}~\bibnamefont{Cichy}},
  \bibinfo{author}{\bibfnamefont{M.}~\bibnamefont{Constantinou}},
  \bibinfo{author}{\bibfnamefont{K.}~\bibnamefont{Hadjiyiannakou}},
  \bibinfo{author}{\bibfnamefont{K.}~\bibnamefont{Jansen}},
  \bibinfo{author}{\bibfnamefont{F.}~\bibnamefont{Steffens}}, \bibnamefont{and}
  \bibinfo{author}{\bibfnamefont{C.}~\bibnamefont{Wiese}},
  \bibinfo{journal}{Phys. Rev.} \textbf{\bibinfo{volume}{D96}},
  \bibinfo{pages}{014513} (\bibinfo{year}{2017}{\natexlab{a}}),
  \eprint{1610.03689}.

\bibitem[{\citenamefont{Ishikawa et~al.}(2017)\citenamefont{Ishikawa, Ma, Qiu,
  and Yoshida}}]{Ishikawa:2017faj}
\bibinfo{author}{\bibfnamefont{T.}~\bibnamefont{Ishikawa}},
  \bibinfo{author}{\bibfnamefont{Y.-Q.} \bibnamefont{Ma}},
  \bibinfo{author}{\bibfnamefont{J.-W.} \bibnamefont{Qiu}}, \bibnamefont{and}
  \bibinfo{author}{\bibfnamefont{S.}~\bibnamefont{Yoshida}},
  \bibinfo{journal}{Phys. Rev.} \textbf{\bibinfo{volume}{D96}},
  \bibinfo{pages}{094019} (\bibinfo{year}{2017}), \eprint{1707.03107}.

\bibitem[{\citenamefont{Alexandrou
  et~al.}(2017{\natexlab{b}})\citenamefont{Alexandrou, Cichy, Constantinou,
  Hadjiyiannakou, Jansen, Panagopoulos, and Steffens}}]{Alexandrou:2017huk}
\bibinfo{author}{\bibfnamefont{C.}~\bibnamefont{Alexandrou}},
  \bibinfo{author}{\bibfnamefont{K.}~\bibnamefont{Cichy}},
  \bibinfo{author}{\bibfnamefont{M.}~\bibnamefont{Constantinou}},
  \bibinfo{author}{\bibfnamefont{K.}~\bibnamefont{Hadjiyiannakou}},
  \bibinfo{author}{\bibfnamefont{K.}~\bibnamefont{Jansen}},
  \bibinfo{author}{\bibfnamefont{H.}~\bibnamefont{Panagopoulos}},
  \bibnamefont{and} \bibinfo{author}{\bibfnamefont{F.}~\bibnamefont{Steffens}},
  \bibinfo{journal}{Nucl. Phys.} \textbf{\bibinfo{volume}{B923}},
  \bibinfo{pages}{394} (\bibinfo{year}{2017}{\natexlab{b}}),
  \eprint{1706.00265}.

\bibitem[{\citenamefont{Radyushkin}(2017{\natexlab{a}})}]{Radyushkin:2017cyf}
\bibinfo{author}{\bibfnamefont{A.~V.} \bibnamefont{Radyushkin}},
  \bibinfo{journal}{Phys. Rev.} \textbf{\bibinfo{volume}{D96}},
  \bibinfo{pages}{034025} (\bibinfo{year}{2017}{\natexlab{a}}),
  \eprint{1705.01488}.

\bibitem[{\citenamefont{Orginos et~al.}(2017)\citenamefont{Orginos, Radyushkin,
  Karpie, and Zafeiropoulos}}]{Orginos:2017kos}
\bibinfo{author}{\bibfnamefont{K.}~\bibnamefont{Orginos}},
  \bibinfo{author}{\bibfnamefont{A.}~\bibnamefont{Radyushkin}},
  \bibinfo{author}{\bibfnamefont{J.}~\bibnamefont{Karpie}}, \bibnamefont{and}
  \bibinfo{author}{\bibfnamefont{S.}~\bibnamefont{Zafeiropoulos}},
  \bibinfo{journal}{Phys. Rev.} \textbf{\bibinfo{volume}{D96}},
  \bibinfo{pages}{094503} (\bibinfo{year}{2017}), \eprint{1706.05373}.

\bibitem[{\citenamefont{Radyushkin}(2017{\natexlab{b}})}]{Radyushkin:2017ffo}
\bibinfo{author}{\bibfnamefont{A.}~\bibnamefont{Radyushkin}},
  \bibinfo{journal}{Phys. Lett.} \textbf{\bibinfo{volume}{B770}},
  \bibinfo{pages}{514} (\bibinfo{year}{2017}{\natexlab{b}}),
  \eprint{1702.01726}.

\bibitem[{\citenamefont{Radyushkin}(2017{\natexlab{c}})}]{Radyushkin:2016hsy}
\bibinfo{author}{\bibfnamefont{A.}~\bibnamefont{Radyushkin}},
  \bibinfo{journal}{Phys. Lett.} \textbf{\bibinfo{volume}{B767}},
  \bibinfo{pages}{314} (\bibinfo{year}{2017}{\natexlab{c}}),
  \eprint{1612.05170}.

\bibitem[{\citenamefont{Ma and Qiu}(2018)}]{Ma:2017pxb}
\bibinfo{author}{\bibfnamefont{Y.-Q.} \bibnamefont{Ma}} \bibnamefont{and}
  \bibinfo{author}{\bibfnamefont{J.-W.} \bibnamefont{Qiu}},
  \bibinfo{journal}{Phys. Rev. Lett.} \textbf{\bibinfo{volume}{120}},
  \bibinfo{pages}{022003} (\bibinfo{year}{2018}), \eprint{1709.03018}.

\bibitem[{\citenamefont{Constantinou and
  Panagopoulos}(2017)}]{Constantinou:2017sej}
\bibinfo{author}{\bibfnamefont{M.}~\bibnamefont{Constantinou}}
  \bibnamefont{and}
  \bibinfo{author}{\bibfnamefont{H.}~\bibnamefont{Panagopoulos}},
  \bibinfo{journal}{Phys. Rev.} \textbf{\bibinfo{volume}{D96}},
  \bibinfo{pages}{054506} (\bibinfo{year}{2017}), \eprint{1705.11193}.

\bibitem[{\citenamefont{Ji}(2014)}]{Ji:2014gla}
\bibinfo{author}{\bibfnamefont{X.}~\bibnamefont{Ji}}, \bibinfo{journal}{Sci.
  China Phys. Mech. Astron.} \textbf{\bibinfo{volume}{57}},
  \bibinfo{pages}{1407} (\bibinfo{year}{2014}), \eprint{1404.6680}.

\bibitem[{\citenamefont{Abdel-Rehim et~al.}(2017)}]{Abdel-Rehim:2015pwa}
\bibinfo{author}{\bibfnamefont{A.}~\bibnamefont{Abdel-Rehim}}
  \bibnamefont{et~al.} (\bibinfo{collaboration}{ETM}), \bibinfo{journal}{Phys.
  Rev.} \textbf{\bibinfo{volume}{D95}}, \bibinfo{pages}{094515}
  (\bibinfo{year}{2017}), \eprint{1507.05068}.

\bibitem[{\citenamefont{Iwasaki}(1983)}]{Iwasaki:2011np}
\bibinfo{author}{\bibfnamefont{Y.}~\bibnamefont{Iwasaki}}
  (\bibinfo{year}{1983}), \eprint{1111.7054}.

\bibitem[{\citenamefont{Abdel-Rehim et~al.}(2014)}]{Abdel-Rehim:2013yaa}
\bibinfo{author}{\bibfnamefont{A.}~\bibnamefont{Abdel-Rehim}}
  \bibnamefont{et~al.}, \bibinfo{journal}{PoS}
  \textbf{\bibinfo{volume}{LATTICE2013}}, \bibinfo{pages}{264}
  (\bibinfo{year}{2014}), \eprint{1311.4522}.

\bibitem[{\citenamefont{Frezzotti and Rossi}(2004)}]{Frezzotti:2003ni}
\bibinfo{author}{\bibfnamefont{R.}~\bibnamefont{Frezzotti}} \bibnamefont{and}
  \bibinfo{author}{\bibfnamefont{G.~C.} \bibnamefont{Rossi}},
  \bibinfo{journal}{JHEP} \textbf{\bibinfo{volume}{08}}, \bibinfo{pages}{007}
  (\bibinfo{year}{2004}), \eprint{hep-lat/0306014}.

\bibitem[{\citenamefont{Sheikholeslami and
  Wohlert}(1985)}]{Sheikholeslami:1985ij}
\bibinfo{author}{\bibfnamefont{B.}~\bibnamefont{Sheikholeslami}}
  \bibnamefont{and} \bibinfo{author}{\bibfnamefont{R.}~\bibnamefont{Wohlert}},
  \bibinfo{journal}{Nucl. Phys.} \textbf{\bibinfo{volume}{B259}},
  \bibinfo{pages}{572} (\bibinfo{year}{1985}).

\bibitem[{\citenamefont{Alexandrou and Kallidonis}(2017)}]{Alexandrou:2017xwd}
\bibinfo{author}{\bibfnamefont{C.}~\bibnamefont{Alexandrou}} \bibnamefont{and}
  \bibinfo{author}{\bibfnamefont{C.}~\bibnamefont{Kallidonis}},
  \bibinfo{journal}{Phys. Rev.} \textbf{\bibinfo{volume}{D96}},
  \bibinfo{pages}{034511} (\bibinfo{year}{2017}), \eprint{1704.02647}.

\bibitem[{\citenamefont{Bali et~al.}(2016)\citenamefont{Bali, Lang, Musch, and
  Schäfer}}]{Bali:2016lva}
\bibinfo{author}{\bibfnamefont{G.~S.} \bibnamefont{Bali}},
  \bibinfo{author}{\bibfnamefont{B.}~\bibnamefont{Lang}},
  \bibinfo{author}{\bibfnamefont{B.~U.} \bibnamefont{Musch}}, \bibnamefont{and}
  \bibinfo{author}{\bibfnamefont{A.}~\bibnamefont{Schäfer}},
  \bibinfo{journal}{Phys. Rev.} \textbf{\bibinfo{volume}{D93}},
  \bibinfo{pages}{094515} (\bibinfo{year}{2016}), \eprint{1602.05525}.

\bibitem[{\citenamefont{Morningstar and Peardon}(2004)}]{Morningstar:2003gk}
\bibinfo{author}{\bibfnamefont{C.}~\bibnamefont{Morningstar}} \bibnamefont{and}
  \bibinfo{author}{\bibfnamefont{M.~J.} \bibnamefont{Peardon}},
  \bibinfo{journal}{Phys. Rev.} \textbf{\bibinfo{volume}{D69}},
  \bibinfo{pages}{054501} (\bibinfo{year}{2004}), \eprint{hep-lat/0311018}.

\bibitem[{\citenamefont{Chen et~al.}(2018)\citenamefont{Chen, Ishikawa, Jin,
  Lin, Yang, Zhang, and Zhao}}]{Chen:2017mzz}
\bibinfo{author}{\bibfnamefont{J.-W.} \bibnamefont{Chen}},
  \bibinfo{author}{\bibfnamefont{T.}~\bibnamefont{Ishikawa}},
  \bibinfo{author}{\bibfnamefont{L.}~\bibnamefont{Jin}},
  \bibinfo{author}{\bibfnamefont{H.-W.} \bibnamefont{Lin}},
  \bibinfo{author}{\bibfnamefont{Y.-B.} \bibnamefont{Yang}},
  \bibinfo{author}{\bibfnamefont{J.-H.} \bibnamefont{Zhang}}, \bibnamefont{and}
  \bibinfo{author}{\bibfnamefont{Y.}~\bibnamefont{Zhao}},
  \bibinfo{journal}{Phys. Rev.} \textbf{\bibinfo{volume}{D97}},
  \bibinfo{pages}{014505} (\bibinfo{year}{2018}), \eprint{1706.01295}.

\bibitem[{\citenamefont{Martinelli et~al.}(1995)\citenamefont{Martinelli,
  Pittori, Sachrajda, Testa, and Vladikas}}]{Martinelli:1994ty}
\bibinfo{author}{\bibfnamefont{G.}~\bibnamefont{Martinelli}},
  \bibinfo{author}{\bibfnamefont{C.}~\bibnamefont{Pittori}},
  \bibinfo{author}{\bibfnamefont{C.~T.} \bibnamefont{Sachrajda}},
  \bibinfo{author}{\bibfnamefont{M.}~\bibnamefont{Testa}}, \bibnamefont{and}
  \bibinfo{author}{\bibfnamefont{A.}~\bibnamefont{Vladikas}},
  \bibinfo{journal}{Nucl. Phys.} \textbf{\bibinfo{volume}{B445}},
  \bibinfo{pages}{81} (\bibinfo{year}{1995}), \eprint{hep-lat/9411010}.

\bibitem[{\citenamefont{Alexandrou
  et~al.}(2017{\natexlab{c}})\citenamefont{Alexandrou, Constantinou, and
  Panagopoulos}}]{Alexandrou:2015sea}
\bibinfo{author}{\bibfnamefont{C.}~\bibnamefont{Alexandrou}},
  \bibinfo{author}{\bibfnamefont{M.}~\bibnamefont{Constantinou}},
  \bibnamefont{and}
  \bibinfo{author}{\bibfnamefont{H.}~\bibnamefont{Panagopoulos}}
  (\bibinfo{collaboration}{ETM}), \bibinfo{journal}{Phys. Rev.}
  \textbf{\bibinfo{volume}{D95}}, \bibinfo{pages}{034505}
  (\bibinfo{year}{2017}{\natexlab{c}}), \eprint{1509.00213}.

\bibitem[{\citenamefont{Accardi et~al.}(2016)\citenamefont{Accardi, Brady,
  Melnitchouk, Owens, and Sato}}]{Accardi:2016qay}
\bibinfo{author}{\bibfnamefont{A.}~\bibnamefont{Accardi}},
  \bibinfo{author}{\bibfnamefont{L.~T.} \bibnamefont{Brady}},
  \bibinfo{author}{\bibfnamefont{W.}~\bibnamefont{Melnitchouk}},
  \bibinfo{author}{\bibfnamefont{J.~F.} \bibnamefont{Owens}}, \bibnamefont{and}
  \bibinfo{author}{\bibfnamefont{N.}~\bibnamefont{Sato}},
  \bibinfo{journal}{Phys. Rev.} \textbf{\bibinfo{volume}{D93}},
  \bibinfo{pages}{114017} (\bibinfo{year}{2016}), \eprint{1602.03154}.

\bibitem[{\citenamefont{Alekhin et~al.}(2017)\citenamefont{Alekhin, Blümlein,
  Moch, and Placakyte}}]{Alekhin:2017kpj}
\bibinfo{author}{\bibfnamefont{S.}~\bibnamefont{Alekhin}},
  \bibinfo{author}{\bibfnamefont{J.}~\bibnamefont{Blümlein}},
  \bibinfo{author}{\bibfnamefont{S.}~\bibnamefont{Moch}}, \bibnamefont{and}
  \bibinfo{author}{\bibfnamefont{R.}~\bibnamefont{Placakyte}},
  \bibinfo{journal}{Phys. Rev.} \textbf{\bibinfo{volume}{D96}},
  \bibinfo{pages}{014011} (\bibinfo{year}{2017}), \eprint{1701.05838}.

\bibitem[{\citenamefont{Ball et~al.}(2017)}]{Ball:2017nwa}
\bibinfo{author}{\bibfnamefont{R.~D.} \bibnamefont{Ball}} \bibnamefont{et~al.}
  (\bibinfo{collaboration}{NNPDF}), \bibinfo{journal}{Eur. Phys. J.}
  \textbf{\bibinfo{volume}{C77}}, \bibinfo{pages}{663} (\bibinfo{year}{2017}),
  \eprint{1706.00428}.

\bibitem[{\citenamefont{de~Florian et~al.}(2009)\citenamefont{de~Florian,
  Sassot, Stratmann, and Vogelsang}}]{deFlorian:2009vb}
\bibinfo{author}{\bibfnamefont{D.}~\bibnamefont{de~Florian}},
  \bibinfo{author}{\bibfnamefont{R.}~\bibnamefont{Sassot}},
  \bibinfo{author}{\bibfnamefont{M.}~\bibnamefont{Stratmann}},
  \bibnamefont{and}
  \bibinfo{author}{\bibfnamefont{W.}~\bibnamefont{Vogelsang}},
  \bibinfo{journal}{Phys. Rev.} \textbf{\bibinfo{volume}{D80}},
  \bibinfo{pages}{034030} (\bibinfo{year}{2009}), \eprint{0904.3821}.

\bibitem[{\citenamefont{Ethier et~al.}(2017)\citenamefont{Ethier, Sato, and
  Melnitchouk}}]{Ethier:2017zbq}
\bibinfo{author}{\bibfnamefont{J.~J.} \bibnamefont{Ethier}},
  \bibinfo{author}{\bibfnamefont{N.}~\bibnamefont{Sato}}, \bibnamefont{and}
  \bibinfo{author}{\bibfnamefont{W.}~\bibnamefont{Melnitchouk}},
  \bibinfo{journal}{Phys. Rev. Lett.} \textbf{\bibinfo{volume}{119}},
  \bibinfo{pages}{132001} (\bibinfo{year}{2017}), \eprint{1705.05889}.

\end{thebibliography}

\end{document}